\documentclass[reprint, aps,prd,twocolumn,showpacs,showkeys, 10pt, longbibliography, nolinenumbers, floatfix]{revtex4-2}
\usepackage{graphicx}
\usepackage{dcolumn}
\usepackage{bm}
\usepackage{amsmath}
\usepackage{amsfonts}

\usepackage[colorlinks = true,linkcolor = blue,urlcolor  = blue,citecolor = blue,anchorcolor = blue]{hyperref}
\usepackage{float}
\usepackage{times}
\usepackage{orcidlink}
\usepackage{xcolor}
\usepackage{multirow}
\usepackage{braket}
\usepackage[normalem]{ulem}

\usepackage[
  protrusion=true,
  expansion=true,
  tracking=true,
  kerning=true,
  spacing=true
]{microtype}

\begin{document}

\title{Hadron-quark phase transitions along proto-neutron star evolution}

\author{P. Laskos-Patkos\orcidlink{0000-0001-5388-2818}$^{1}$}
\email{plaskos@physics.auth.gr}
\author{P.S. Koliogiannis\orcidlink{0000-0001-9326-7481}$^{2}$}
\email{pkoliogi@phy.hr}
\author{Ch.C. Moustakidis\orcidlink{0000-0003-3380-5131}$^{1}$}
\email{moustaki@auth.gr}

\affiliation{$^1$Department of Theoretical Physics, Aristotle University of Thessaloniki, 54124 Thessaloniki, Greece}
\affiliation{$^2$Department of Physics, Faculty of Science, University of Zagreb, Bijeni{\v c}ka cesta 32, 10000, Zagreb, Croatia.}

\begin{abstract}
The new era of multi-messenger astronomy requires the accurate and self-consistent derivation of the nuclear equation of state at high temperature. In the present work, we focused on the calculation of hot hybrid equations of state, studying the different evolution stages of a proto-neutron star with a quark matter core (proto-hybrid star). For the hadronic matter we used two distinct Skyrme effective interactions, while for quark matter the well-known vector MIT bag model was employed. To model the era of trapped neutrinos in the system we considered the global conservation of lepton fraction which resulted in an equation of state with an extended mixed phase. For periods following the neutrino diffusion phase of a proto-neutron star, the equations of state were modelled using both the Maxwell  and the Gibbs construction depending on the assumption for either local or global electric-charge conservation. With the use of the derived hybrid models, we solved the Tolman-Oppenheimer-Volkov equations to describe the corresponding hybrid star configurations. Finally, we investigated how the structure of proto-hybrid stars evolves, using constant rest mass sequences. We found that regardless of whether electric-charge is globally or locally conserved, the earlier stages of a hybrid star's life may play a crucial role on the determination of its maximum possible gravitational mass in later stages.

\end{abstract}
\maketitle

\section{Introduction}

Neutron stars are often considered to be unique extraterrestrial laboratories for the study of dense and strongly-interacting matter~\cite{Bielich-2020,Glendenning-2000,Haensel2007NeutronStars1,ShapiroTeukolsky1983,Lattimer-2001,Lattimer-2004,Koliogiannis_2021_rev}. In particular, while the subject of compact star physics has been a theoretical endeavour for decades,  recent simultaneous measurements of neutron-star masses and radii~\cite{Cromartie-2020,Miller-2019,Riley-2019,Salmi-2024a,Choudhury-2024,Salmi-2024,Vinciguerra-2024} have provided direct astrophysical constraints on the neutron-star equation of state (EOS), complementary to those inferred from finite nuclei and nuclear experiments at low densities~\cite{Lattimer_2021,Raaijmakers-2019}. In addition, the discovery of gravitational waves (GWs) and the detection of such radiation from the inspiral phase of binary neutron star mergers has also yielded important information on other bulk stellar properties (e.g., tidal deformability), setting the ground for further constraining the properties of nuclear matter~\cite{Abbott-2017,Abbott-2018}. Notably, neutron-star measurements are being used not only to constrain the EOS-namely, the relation between pressure and energy density-but also to potentially shed light on the microphysics of neutron star cores. More precisely, several intriguing hypotheses (along with their implications) such as the existence of deconfined quark matter or boson condensates, can now be put to the test against empirical data~\cite{Nandi-2018,Montana-2019,Annala-2020,Annala-2023,Laskos-2024,Li-2024,Sharifi-2021,Veselsky-2025}. 

Interestingly, the vast majority of neutron star studies do not incorporate the effects of temperature on the nuclear EOS~\cite{Bielich-2020,Glendenning-2000}. Such an approach is well justified  when  modelling the interiors of old neutron stars. Despite the fact that the actual temperature in their core might be close to $0.1$ MeV ( $\sim 10^9$ K), its contribution to the energy density and pressure is negligible due to the fact that matter is extremely degenerate~\cite{Bielich-2020,Glendenning-2000}. In contrast, during the formation and early evolution of compact objects, such as in core-collapse supernovae and binary neutron-star merger remnants, temperatures may become sufficiently high for thermal effects to be significant~\cite{Prakash-1997,Lattimer-1981,Baiotti-2017,Radice-2020}. Given that advances in GW astronomy may now allow us to have accurate measurements about neutron stars during their formation (e.g., gravitational waves related to non-radial oscillation modes)~\cite{Abdikamalov-2021,Radice-2019,DePietri-2020}, it  is evident that the accurate and self-consistent determination of the nuclear EOS at finite temperature is of particular importance. Over the years, considerable attention has been devoted to investigating the effects of temperature on the properties of dense nuclear matter and proto-neutron stars~\cite{Lamb-1978,Bethe-1979,Brown-1982,Lattimer-1991,MoustakidisPanos2009,Moustakidis2009,Burgio-2010,Wellenhofer-2015,Lu-2019,Li-2021,Constantinou-2014,Constantinou-2015,Koliogiannis-2021,Sedrakian-2021,daSilva-2025,Thakur-2025,Drago-1999,Pagliara-2010,Roark-2018,Malfatti-2019,Roark-2019,Malfatti-2019,Logoteta-2022,KhosraviLargani-2022,Carlomagno-2024,Carlomagno-2024-twin,Ghaemmaghami-2025,Sabatucci-2026,kunkel-2026,Steiner-2000,Hempel-2009,Issifu-2025,Pagliara-2009,Mariani-2017}.  Earlier works  focused mainly on phenomenological approaches~\cite{Lamb-1978,Bethe-1979,Brown-1982,Lattimer-1991}, while later several studies worked on microscopic calculations at finite temperature~\cite{Burgio-2010,Wellenhofer-2015,Lu-2019,Li-2021}. In addition, numerous studies have investigated the presence of exotic forms of matter, such as hyperons or other heavy baryons~\cite{Sedrakian-2021,daSilva-2025,Thakur-2025}, and deconfined quarks~\cite{Drago-1999,Pagliara-2010,Roark-2018,Malfatti-2019,Roark-2019,Malfatti-2019,Logoteta-2022,KhosraviLargani-2022,Carlomagno-2024,Carlomagno-2024-twin,Ghaemmaghami-2025,Sabatucci-2026,kunkel-2026,Steiner-2000,Hempel-2009,Issifu-2025,Pagliara-2009,Mariani-2017}, in the EOS.

At a first approximation hot EOSs are studied in an isothermal framework, meaning that temperature is considered to be uniform throughout the star~\cite{Constantinou-2014,Constantinou-2015,Koliogiannis-2021}. A more advanced approach that is considered to provide a more accurate description for the early evolutionary stages of a neutron star is the isentropic approximation~\cite{Prakash-1997}. In such a scenario, the constant quantity is not the temperature but the entropy per baryon. As mentioned in Ref.~\cite{Prakash-1997}, which studied the early stages of a proto-neutron star's life, the entropy per baryon reaches values of the order $1-2$~$k_B$. In addition, due to the fact that newly-born compact stars are very opaque, neutrinos are trapped in hot matter, significantly increasing the number of leptons in the system~\cite{Prakash-1997,Steiner-2000}. 

As previously mentioned, apart from studying the case of purely hadronic compact stars, a lot of studies have investigated the importance of temperature on the properties of proto-neutron stars with a quark matter core (proto-hybrid stars)~\cite{Drago-1999,Pagliara-2010,Roark-2018,Malfatti-2019,Roark-2019,Malfatti-2019,Logoteta-2022,KhosraviLargani-2022,Carlomagno-2024,Carlomagno-2024-twin,Ghaemmaghami-2025,Sabatucci-2026,kunkel-2026,Steiner-2000,Hempel-2009,Issifu-2025,Pagliara-2009,Mariani-2017}. In particular, the description of hadron-quark phase transition at finite temperature is of significant interest since, based on the assumptions concerning the globally and locally conserved charges in the system, one can get  different viable configurations for proto-hybrid stars~\cite{Roark-2018,Roark-2019}. For instance, the global conservation of the electric charge or the lepton fraction, would lead to a configuration with an extended mixed phase containing both hadrons and quarks~\cite{Glendenning-1992}. In contrast, imposing the local conservation of those charges yields a hybrid EOS that is characterized by a sharp interface between the two phases, accompanied by an energy-density discontinuity~\cite{Bielich-2020}.

The construction of a hot hybrid EOS with a density discontinuity (local charge neutrality assumption) is not always a trivial task~\cite{Hempel-2009}. When studying cold hybrid stars, one needs  to simply equate the baryon chemical potential and the pressure for both phases, marking the onset of phase transition. This approach is also directly applicable to the finite-temperature extension of a hybrid EOS  within the isothermal framework~\cite{Hempel-2009}. Following that, several studies proceed in a similar manner for the construction of hot hybrid models within the isentropic framework, which is a reasonable first approximation~\cite{Mariani-2017,Roark-2018,Roark-2019,Carlomagno-2024-twin,Issifu-2025}. However, it should be noted that the consideration of a fixed and equal entropy per baryon in both phases, combined with the enforcement of equal pressure and baryon chemical potential, will inevitably lead to a discontinuity in  temperature, violating the thermal equilibrium condition between the two phases~\cite{Roark-2018}. A possible solution for that issue involves the construction a mixed phase  interpolating between the two isentropic EOSs (hadronic and quark), where the entropy per baryon is treated as a conserved thermodynamic potential~\cite{kunkel-2026}. Another approach would be that of Ref.~\cite{Malfatti-2019}, where the authors proceeded by selecting a starting transition point on the phase diagram and constructing the relevant isentropic EOSs. In that work, the resulting compact star models turned unstable either immediately or shortly after the central pressure reached the phase transition onset, suggesting the absence of deconfined quark matter in proto-neutron stars. By contrast, this temperature discontinuity is non-present in systems where  the global conservation of any additional charge (apart from the baryon number) is considered. In neutrino-free matter, this can be achieved through global electric-charge conservation~\cite{Drago-1999,Logoteta-2022}, whereas in neutrino-trapped matter, global conservation of the lepton number provides the additional degree of freedom required to satisfy thermal equilibrium~\cite{Pagliara-2009,Pagliara-2010,Sabatucci-2026}.

The main goal of the present study is to provide a self-consistent description of the EOSs describing each stage of a proto-hybrid star. In particular, under the consideration of global lepton number conservation, we study the evolution of proto-hybrid stars in two distinct scenarios: (a) local electric-charge conservation (LCN), and (b) global electric-charge conservation (GCN). To investigate the hadron-quark phase transition and assess the influence of the hadronic EOS stiffness on its properties, we employ two hadronic models based on Skyrme interactions~\cite{Skyrme-1958,Kohler-1976,Reinhard-1995}, together with a quark-matter EOS based on the MIT bag model~\cite{Klahn-2015,Lopes-2021a,Lopes-2021b,Gomes-2019}. To tackle the issues related to the temperature discontinuity in hybrid models with sharp phase transitions (LCN case), we allow the two phases to have different entropies per baryon and evaluate both the resulting entropy per baryon discontinuity and its impact on the bulk properties of proto-hybrid stars.  Finally, we examine the impact of the EOS construction scheme (LCN and GCN) on the evolution of proto-hybrid stars, by constructing the relevant rest mass-gravitational mass diagrams and discussing the corresponding constant rest mass sequences.

This paper is organized as follows. In Sec.~\ref{section2} we present the model considered for the finite temperature description of hadronic matter, while in Sec.~\ref{section3} the relevant quark model is given. Sec.~\ref{section4} contains some analytical approximations that we employed in order to test the validity of our numerical calculations. In Sec.~\ref{section5} we discuss the stages of a proto-neutron star evolution paying special attention to the calculation recipe for the construction of each hybrid EOS. In Sec.~\ref{section6} we present our results and their implications. Finally, Sec.~\ref{section7} contains a summary of our findings. Throughout this study, we employ natural units, $\hbar=c=k_{B}=1$.

\section{Hadronic Matter} \label{section2}

For the description of the hadronic phase we will rely on the well-known Skyrme~\cite{Skyrme-1958} Hamiltonian density (or energy density) which is given as~\cite{Constantinou-2014}
\begin{equation}\label{eq1}
\begin{split}
\mathcal{H} & =
\frac{\tau_n}{2m_n}+\frac{\tau_p}{2m_p}
\\
&+
n_b(\tau_n+\tau_p)\left[\frac{t_1}{4}\left(1+\frac{x_1}{2}\right)+\frac{t_2}{4}\left(1+\frac{x_2}{2}\right)\right]
\\
&+
(\tau_nn_n+\tau_pn_p)\left[\frac{t_2}{4}\left(\frac{1}{2}+x_2\right)-\frac{t_1}{4}\left(\frac{1}{2}+x_1\right)\right]
\\
&+
\frac{t_0}{2}\left(1+\frac{x_0}{2}\right)n_b^2-\frac{t_0}{2}\left(\frac{1}{2}+x_0\right)(n_n^2+n_p^2)
\\
&+
\left[\frac{t_3}{12}\left(1+\frac{x_3}{2}\right)n_b^2-\frac{t_3}{12}\left(\frac{1}{2}+x_3\right)(n_n^2+n_p^2)\right]n_b^\alpha.
\end{split}
\end{equation}
Here, $t_j$ ($j=0,\ldots,3$), $x_j$ ($j=0,\ldots,3$),  and $\alpha$ are the Skyrme interaction parameters, $n_i$ ($i=n,p$) denote the nucleon number densities, $\tau_i$ are the corresponding kinetic energy densities and $n_b=n_n+n_p$ is the baryon density. 
The number density of  neutrons or protons is given by
\begin{equation} \label{e2}
    n_i = g \int \frac{d^3k}{(2\pi)^3}f_i(n_n,n_p,k,T),
\end{equation}
where $g$ is the degeneracy factor.
The kinetic energy terms $\tau_i$ read as
\begin{equation}\label{eq3}
   \tau_i=
    g \int \frac{d^3k}{(2\pi)^3} k^2f_i(n_n,n_p,k,T),
\end{equation}
while $f_i(n_n,n_p,k,T)$ corresponds to the Fermi-Dirac distribution
\begin{equation}\label{eq4}
    f_i=\left\{1+\exp\left[\frac{e_i(n_n,n_p,k,T)-\mu_i(n_n,n_p,T)}{T}\right]\right\}^{-1},
\end{equation}
with the quantities $e_i$ and $\mu_i$ denoting the single-particle energy and chemical potential for $i$ particles. The degeneracy parameter $g$ is equal to $2$ for neutrons and protons.

The single particle-energy can be evaluated through functional derivatives of the Hamiltonian density~\cite{Constantinou-2014}
\begin{equation}\label{eq5}
    e_i=k^2\frac{\partial \mathcal{H}}{\partial \tau_i}+\frac{\partial \mathcal{H}}{\partial n_i},
\end{equation}
and is interpreted as the energy spectrum of a quasi-particle moving in the mean-field generated by the surrounding medium. In that sense, after evaluating Eq.~\eqref{eq5}, and given that Skyrme models are non-relativistic, one should be able to rewrite this quantity as
\begin{equation}\label{eq6}
    e_i=\frac{k^2}{2m} + U_i(n_n,n_p,k),
\end{equation}
where $U_i$ is known as the single-particle potential. Notably, for traditional Skyrme models the single-particle potential has a dependence on momentum that is of the form $U_i\sim k^2$~\cite{Costantinou-2015b}. 

The determination of the hadronic EOS at finite temperature or entropy requires the imposition of two additional constraints related to chemical equilibrium and charge neutrality~\cite{Bielich-2020}. The exact form of these constraints depends on the considered composition of the system. In the present work, we will be considering nuclear matter systems containing neutrons, protons and electrons. In addition, for the modelling of the early proto-neutron star evolution stages the presence of neutrinos will also be incorporated. For a system containing neutrons ($n$), protons ($p$) and electrons ($e$) the requirement for chemical equilibrium provides the following relation for the chemical potentials of different particles~\cite{Prakash-1997}
\begin{equation} \label{eq7}
\mu_n=\mu_p+\mu_e,
\end{equation}
while for a system that also contains neutrinos ($\nu$) we have~\cite{Prakash-1997}
\begin{equation}\label{eq8}
\mu_n+\mu_\nu=\mu_p+\mu_e.
\end{equation}
In both cases the charge neutrality demand is formulated by equating the number of protons and electrons which are the electrically charged particles of the systems. Therefore we have~\cite{Bielich-2020}
\begin{equation}\label{eq9}
    n_p=n_e.
\end{equation}

With the use of the formalism presented above one can evaluate the energy density $\mathcal{E}_H=\mathcal{H}$ of hadronic matter for a given baryon density and temperature. Furthermore, the entropy density related to each particle species can be evaluated with the use of the following formula
\begin{equation}\label{eq10}
     s_i=-g\int \frac{d^3k}{(2\pi)^3}[f_i \ln(f_i)+(1-f_i)\ln(1-f_i)].
\end{equation}
With knowledge of all of these quantities one can evaluate the hadronic pressure via the standard thermodynamic relation 
\begin{equation}\label{eq11}
    P_{H}=-\mathcal{E}_H+Ts+\sum_{i=n,p}\mu_in_i.
\end{equation}

The resulting finite-temperature EOS requires the consideration of lepton contributions to the energy density and pressure. Since leptons are treated as a relativistic Fermi gas, the aforementioned quantities are given by
\begin{equation}\label{eq12}
    \mathcal{E}_l
=
\frac{g}{2\pi^2}
\int_0^\infty
k^2 \, \sqrt{k^2 + m_l^2}
f_l
\, dk,
\end{equation}
\begin{equation}\label{eq13}
P_l = \frac{g}{6 \pi^2} \int_0^\infty \frac{k^4}{\sqrt{k^2 + m_l^2}} 
f_l \, dk,
\end{equation}
where $f_l$ is the Fermi-Dirac distribution for leptons given as
\begin{equation}\label{eq14}
    f_l=\frac{1}{\exp\left[\frac{\sqrt{k^2 + m_l^2} - \mu_l}{T}\right] + 1}.
\end{equation} 
The degeneracy parameter $g$ takes the value 2 for electrons and 1 for neutrinos. Note that Eq.~\eqref{eq10} can also be used to evaluate the entropy density of leptons. Then, the total energy density, pressure, and entropy density of the system will be given as
\begin{equation}\label{eq15}
    \mathcal{E}_{tot}=\mathcal{E}_H+\sum_{l}\mathcal{E}_l,
\end{equation}
\begin{equation}\label{eq16}
    P_{tot}=P_H+\sum_{l}P_l,
\end{equation}
\begin{equation}\label{eq17}
    s_{tot}=\sum_{i=n,p}s_i+\sum_{l}s_l.
\end{equation}

Finally, a note is appropriate with regard to the numerical approach used for the determination of the hot hadronic EOS. In present work we have closely followed the formalism presented in Ref.~\cite{Constantinou-2014} which proceeds to finite temperature calculations with the use of accurate interpolations of non relativistic Fermi-Dirac integrals. The employed interpolations have been presented in Ref.~\cite{Cody-1967}.

\section{Quark Matter} \label{section3}

For the description of the quark phase we used a widely employed modification of the MIT bag model~\cite{Klahn-2015,Lopes-2021a,Lopes-2021b,Gomes-2019,Gomes-2019b,Jaikumar-2021,Constantinou-2021,Constantinou-2022,Constantinou-2022b}. The standard MIT bag model was introduced as an effective way to describe the structure of baryons. In this approach, quarks are confined and  allowed to move freely within a 'bag', which they cannot escape due to the presence of an external pressure (bag constant/pressure). The corresponding Lagrangian density of the standard MIT bag model reads as~\cite{Lopes-2021a}
\begin{equation} \label{eq18}
    \mathcal{L}_0=\left\{\sum_{q}[\bar{\psi_q} (i\gamma_{\mu}\partial^{\mu}-m_q)\psi_q-B\right\}\Theta,
\end{equation}
where $\psi_q$ and $m_q$ denote the Dirac field and the mass of $q$ quarks, respectively, and $B$ is the so-called bag constant. $\Theta$ is a step function that takes the value zero outside of the bag, ensuring the confinement.

The consideration of the aforementioned Lagrangian density would result into a particularly soft EOS which would most likely fail to account for the existence of massive compact stars (above $2M_\odot$). This is clear by the fact that the only positive contributions to the pressure result from the degenarate nature of the Fermionic system. In order to tackle such issues, we consider the presence of a vector boson that mediates a repulsive force among quarks. The corresponding term in the Lagrangian density would then be given by~\cite{Lopes-2021a}
\begin{equation} \label{eq20}
    \mathcal{L}_{\rm vec}=\left\{-g_v\sum_{q} \bar{\psi_q} \gamma_\mu V^\mu\psi_q + \frac{1}{2}m_{V}^2 V_\mu V^\mu\right\}\Theta,
\end{equation}
where $g_v$ is the coupling constant of the interaction and $m_{V}$ is the boson mass.

Using the mean-field approximation one can derive the single-particle energy of $q$ quarks as~\cite{Lopes-2021a}
\begin{equation}\label{eq21}
    e_q=\sqrt{k^2+m_q^2}+g_v V^0,
\end{equation}
\begin{equation}\label{eq22}
    m_vV_0=\sum_{q=u,d,s}g_vn_q,
\end{equation}
where $n_q=\braket{\bar{\psi}_q\gamma^0\psi_q}$ is the q quark number density. With knowledge of this single-particle energy one can evaluate the full thermodynamic state of the system at finite temperature. In particular,
the number density, energy density, and pressure of q quarks will be given by~\cite{Lopes-2021b}
\begin{equation} \label{eq23}
    n_q = g \int \frac{d^3k}{(2\pi)^3}f_q,
\end{equation}
\begin{equation}\label{eq24}
    \mathcal{E}_Q=\sum_{q=u,d,s}\mathcal{E}_q-\braket{\mathcal{L}},
\end{equation}
\begin{equation}\label{eq25}
    P_Q=\sum_{q=u,d,s}P_q+\braket{\mathcal{L}},
\end{equation}
where $\mathcal{E}_q$ and $P_q$ are calculated with the use of standard Fermi-Dirac statistics as~\cite{Lopes-2021b}
\begin{equation}\label{eq26}
    \mathcal{E}_q=g \int \frac{d^3k}{(2\pi)^3}e_qf_q,
\end{equation}
\begin{equation}\label{eq27}
    P_q=\frac{g}{3}\int \frac{d^3k}{(2\pi)^3}k\frac{\partial e_q}{\partial k}f_q,
\end{equation}
and $\braket{\mathcal{L}}$ is related to contributions from the bag constant and the vector boson given by~\cite{Lopes-2021b}
\begin{equation}\label{eq28}
    \braket{\mathcal{L}}=-B+\frac{1}{2}m_V^2V_0^2.
\end{equation}
Obviously, $f_q$ stands for the Fermi-Dirac distribution similar to Eq.~\eqref{eq4}.

As in the case of hadronic matter, the calculation of the quark EOS requires the imposition of chemical equilibrium and charge neutrality. Considering the presence of quarks (up, down, strange) and electrons, the demand for chemical equilibrium leads to~\cite{Lopes-2021a,Lopes-2021b}
\begin{equation} \label{eq29a}
    \mu_d=\mu_u+\mu_e, \quad \mu_d=\mu_s,
\end{equation}
while charge neutrality is expressed via
\begin{equation} \label{eq29b}
    \frac{2}{3}n_u-\frac{1}{3}(n_d+n_s)-n_e=0.
\end{equation}
In the presence of neutrinos, Eq.~\eqref{eq29a} is modified as
\begin{equation} \label{eq30}
    \mu_d+\mu_\nu=\mu_u+\mu_e, \quad \mu_d=\mu_s.
\end{equation}
Note that the final EOS will result by summing the contributions of both quarks and leptons to the energy density and pressure, respectively. The leptonic contributions to the EOS have been thoroughly discussed in Sec~\ref{section2}. 

Finally, in the present study we did not consider the presence of anti-quarks in the systems. Such contribution would be modelled by adding in the integrals of Eqs.~\eqref{eq23}~-~\eqref{eq25} another Fermi-Dirac distribution with chemical potential opposite to those of ordinary quarks (and hence negative)~\cite{Lopes-2021b}. For typical neutron star conditions we have performed the relevant calculations and seen that the anti-quark contributions can be safely neglected.

\section{Analytical approximations} \label{section4}

In an interesting work, Constantinou {\it et al.}~\cite{Costantinou-2015b} have managed to calculate next-to-leading order terms for the thermodynamical properties of a degenerate Fermi-Dirac system, extending significantly the range of validity of analytical approximations for the nuclear EOS at finite temperature. In the present work, we employ the framework they provided in order to test that our numerical calculations are correct for both phases. As the hadronic model is a non-relativistic potential model, while the quark phase is described by a relativistic mean-field model, the implementation of the analytical relations from Ref.~\cite{Costantinou-2015b} is slightly different for each case. The main analytical relations are presented in the following subsections for completeness.

\subsection{Sommerfeld expansion for non-relativistic potential models}

In practice, the details of a particular nuclear model are inserted into the Sommerfeld expansion terms via the Landau effective mass (see definition below) and its derivatives with respect to momentum. More precisely, the detailed calculations of Ref.~\cite{Costantinou-2015b} have shown that, at the degenerate limit, the entropy per particle for $i$ particles is given by the following expression
\begin{equation}\label{eq31}
    S_i=2a_iT - \frac{16}{5\pi^2}a_i^3T^3(1 - L_{Fi}),
\end{equation}
where
\begin{equation}\label{eq32}
    a_i=\frac{\pi^2m_i^*}{2k_{Fi}^2}, \hspace{0.3 cm} L_{F_i}=\frac{7}{12}\frac{\mathcal{M}^{'2}_{F_i}}{m^{*2}}+\frac{7}{12}\frac{\mathcal{M}^{''}_{F_i}}{m^{*}}+\frac{3}{4}\frac{\mathcal{M}_{F_i}^{'}}{m^{*}}.
\end{equation}

The Landau effective mass $\mathcal{M}_i$ of $i$ particles is defined as~\cite{Costantinou-2015b}
\begin{equation}\label{eq33}
    \frac{\mathcal{M}_i(n_n.n_p,k)}{m_i}=\left\{1+\frac{m_i}{k}\frac{\partial U_i(n_n,n_p,k)}{\partial k}\right\}^{-1}.
\end{equation}
Then, $\mathcal{M}_{F_i}$ is simply the value of the Landau effective mass at the Fermi surface ($\mathcal{M}_{F_i}=\mathcal{M}_i(n_n,n_p,k=k_F)$). Note that $m_i^*$ denotes $\mathcal{M}_{F_i}$.

Knowing the entropy per particle allows the evaluation of all of the other relevant thermodynamic properties via standard relations (see Ref.~\cite{Costantinou-2015b}). In the present study, we are interested in evaluating the thermal energy per particle and the pressure, which are defined as
\begin{align}
    E_i^{th}&=E_i(n_n,n_p,T)-E_i(n_n,n_p,0), \\
    P_i^{th}&=P_i(n_n,n_p,T)-P_i(n_n,n_p,0).
\end{align}

A final remark is that since the momentum dependence in the single-particle potential of Skyrme models is of the form $U_i(n_n,n_p,k)\sim k^2$, the derivatives of the Landau effective mass vanish. As a consequence, the term $L_{F_i}$ is zero for the hadronic model employed in the present work~\cite{Costantinou-2015b}.

\subsection{Sommerfeld expansion for RMF models}

As in the case of potential models, the Sommerfeld expansion draws information about the characteristics of a model via an effective mass $M^*$. The use of a different symbol is related to the fact that the effective mass has a different definition for relativistic models~\cite{Costantinou-2015b}. In particular, $M^*$ denotes the Dirac effective mass, which is directly related to the presence of scalar interactions in the system. 

For RMF models the entropy per particle will be given once again by Eq.~(\ref{eq31}) with a change in the definition of $a_i$ and $L_{F_i}$~\cite{Costantinou-2015b}. In particular,
\begin{equation}\label{eq34}
     a_i=\frac{\pi^2E_{F_i}^*}{2p_{Fi}^2}, \hspace{0.5 cm} L_{F_i}=\frac{11}{12}\frac{E^{*2}_{F_i}}{p_{Fi}^2}-\frac{5}{12}\frac{E^{*4}_{F_i}}{p_{Fi}^4},
\end{equation}
where $E_{F_i}^*=\sqrt{k_{F_i}^2+M_i^{*2}}$. After the determination of $S_i$ everything else can be derived via standard thermodynamics.

Notably, since in the present study, the quark model incorporates only a vector interaction, the Dirac effective mass is going to coincide with the rest mass for each particle.

\section{Proto-neutron star Evolution stages and phase transition} \label{section5}

In the present study, we will focus on three different EOSs, each characterizing a different stage of proto-neutron star evolution. Below, we briefly review the thermodynamic conditions for each stage following Refs.~\cite{Prakash-1997,Steiner-2000,Mariani-2017}. 

During the first stage, immediately after the formation of a proto-neutron star, the mean-free path of neutrinos (created by beta decay and its inverse process) is small compared to the stellar radius. As a consequence, neutrinos are trapped in the stellar interior. The EOS related to this configuration will be characterized by an entropy per baryon $S_b=1$ and a lepton fraction of $Y_l=0.4$~\cite{Prakash-1997,Steiner-2000,Mariani-2017}.

After $\sim 20$ seconds, the compact star is deleptonized due to neutrino diffusion. As a result, the neutrino density ($n_\nu$) goes to zero. In addition, the star is heated due the fact that neutrino diffusion leaves most of the neutrinos' energy to the stellar core. During this stage, a system is often characterized in the literature by $S_b=2$ and absence of neutrinos~\cite{Prakash-1997,Steiner-2000,Mariani-2017}.

Finally, after a few minutes, the star cools down completely and it will be effectively characterized by zero temperature and hence zero entropy per baryon~\cite{Prakash-1997,Steiner-2000,Mariani-2017}.

    Throughout the following subsections, we thoroughly describe the conditions required for the coexistence of hadronic and quark phases for the aforementioned evolution stages. For reasons of simplicity, we begin with the most trivial case of zero temperature, which is then naturally extended to finite temperature. The three evolution stages will be studied under two distinct assumptions for the system: (a) LCN and (b) GCN.

\subsection{Zero temperature}
\label{sub_secA}

In general, there are two main approaches for the derivation of hybrid EOSs, the Maxwell and the Gibbs construction. At zero temperature, the favoured approach is determined by the currently unknown surface tension ($\sigma$) between the hadronic and the quark phase~\cite{Bielich-2020,Glendenning-1992,Mariani-2017}. If one considers a sufficiently large surface tension, the favoured method is Maxwell construction~\cite{Bielich-2020}. In this approach, each phase satisfies electric charge neutrality locally. In contrast, if the surface tension is low, the favoured method for deriving hybrid EOSs is the Gibbs construction~\cite{Glendenning-1992}.

In the scenario of LCN, the following conditions need to be satisfied~\cite{Bielich-2020}
\begin{equation} \label{eq35}
    P^h=P^q, \hspace{0.5 cm} \mu_b^h=\mu_b^q,  \hspace{0.5 cm} T^h=T^q,
\end{equation}
where $P$ is the pressure, $\mu_b$ is the baryon chemical potential, and $T$ is the temperature. The superscripts $h$ and $q$ denote the two different phases. Note that, since we are discussing the case of cold matter, the requirement for temperature equality is trivially fulfilled. Interestingly, the key characteristic of the Maxwell construction is that the two phases can only coexist for a single value of pressure. Hence, it naturally follows that in the presence of a gravitational field (e.g., in compact stars) the two phases will be separated.

In the case of GCN, the Gibbs construction dictates the following conditions~\cite{Glendenning-1992}
\begin{equation} \label{eq36}
    P^h=P^q, \hspace{0.5 cm} \mu_b^h=\mu_b^q, \hspace{0.5 cm} \mu_Q^h=\mu_Q^q,  \hspace{0.5 cm} T^h=T^q,
\end{equation}
where $\mu_Q$ is the charge chemical potential, which in the context of the present study is equal to the electron chemical potential $\mu_e$. Contrary to the Maxwell construction, the two phases, which are now electrically charged, could coexist for a wide range of pressure~\cite{Glendenning-1992,Hempel-2009}. As a consequence, in the presence of a gravitational field, the phase transition occurs through an extended mixed-phase region, where hadronic and quark matter coexist rather than being separated by a sharp interface. A key quantity for the mathematical modelling of such a mixed phase is the so-called volume fraction ($\chi$) of each phase, which may be derived by enforcing the demand for charge neutrality in the whole system~\cite{Glendenning-1992}
\begin{equation} \label{eq37}
    (1-\chi)n_Q^h+\chi n_Q^q=0.
\end{equation}
By substituting the charge density $n_Q$ for each phase,  Eq.~\eqref{eq37} can be re-written as 
\begin{equation}
    (1-\chi)n_p+\frac{1}{3}\chi \left(2n_u-n_d-n_s\right)-n_e=0.
\end{equation}
As is evident, $\chi$ takes values from 0 to 1, with $\chi=0$ corresponding to a purely hadronic phase and $\chi=1$ dictating the absence of hadrons.

\subsection{Finite temperature and neutrino absence} 

During the second evolution stage, the proto-neutron star is hot and often characterized by constant entropy per baryon equal to $S_b=2$~\cite{Prakash-1997}. In addition, neutrinos are absent so both the relevant degrees of freedom and the conditions for phase equilibrium are the same as in the cold case described in Sec.~\ref{sub_secA}. Once again, we will proceed by considering the two different construction schemes related to assumption on whether electric charge is locally or globally conserved. 

Let us begin with the case of LCN. In this case, the conditions for phase coexistence are given by Eq.~\eqref{eq35}. Note that in the case of zero temperature, the thermal equilibrium demand was trivially satisfied. However, in the case of a hot hybrid star, particular care must be taken in how this condition is implemented. More precisely, the potential enforcement of equal entropy per baryon in both phases (in order to construct an isentropic EOS), would certainly yield a temperature discontinuity at the point of coexistence~\cite{Hempel-2009,Roark-2018}. 

In the present work, for the LCN case, we will proceed by allowing the two phases to have different entropy per baryon in order to ensure that their temperatures equate at the point of coexistence. In particular, to derive the relevant hybrid EOS we will use the following procedure:
\begin{enumerate}
    \item Construction of the phase diagram using Eq.~(\ref{eq35}).
    \item Construction of a hadronic EOS with constant entropy per baryon ($S_b=2$).
    \item Calculation of the intersection between the phase diagram and the derived EOS.
    \item Construction of a quark EOS with constant entropy that intersects the point derived in the previous step.
\end{enumerate}
A similar approach has been used in the work of Ref.~\cite{Malfatti-2019}, where the authors constructed EOSs by selecting the transition temperature (i.e., the fixed temperature at which the transition occurs). 

It is noteworthy that there is a simple way to test the thermodynamic consistency of the derived hybrid EOS. We refer of course to the standard Claussius-Clapeyron equation which provides the slope of the phase diagram curve in the $P-T$ plane. More precisely, the equation reads
\begin{equation}\label{cceq}
    \frac{dP}{dT}=\frac{\Delta S}{\Delta V},
\end{equation}
where $S$ is the total entropy for each phase and $V$ its volume. Rewriting this equation in a form that is more familiar to the quantities found in neutron star physics one gets~\cite{Roark-2018}
\begin{equation}\label{cceq2}
    \frac{dP}{dT}=\frac{S_b^q-S_b^h}{\frac{1}{n_b^q}-\frac{1}{n_b^h}}.
\end{equation}
Note that, the derivation of the hybrid EOS yields the right hand side of Eq.~(\ref{cceq2}) and one needs to verify that the numerical evaluation for the slope $dP/dT$ also leads to the same value. 

The aforementioned issue, related to the temperature discontinuity, is absent system characterized by GCN. In this case, one can proceed with the construction of the hybrid EOS by utilizing Eqs.~\eqref{eq36} and~\eqref{eq37} and enforcing a constant entropy per baryon throughout the system, which will be written as
\begin{equation}
    S_b=\frac{1}{n_b}[(1-\chi)s_b^h+\chi s_b^q],
\end{equation}
with 
\begin{equation}
    n_b=(1-\chi)n_b^h+\chi n_b^q.
\end{equation}
Note, that the entropies per baryon in each phase ($S^h_b,S^q_b$) will once again be different. It is only the entropy per baryon of the mixed phase that can be fixed to be constant.

\subsection{Finite temperature and neutrino trapping}

In comparison to the previously described cases, neutrino trapping significantly alters the phase transition properties in a proto-neutron star. As previously mentioned, for the initial proto-neutron star state the conditions $S_b=1$ and $Y_l=0.4$ are imposed. Following the detailed discussion of Ref.~\cite{Hempel-2009}, there is no physical reason for locally fixed lepton fraction in the two phases as there is no long range force associated with this charge. As a consequence, one can only demand global conservation of the lepton fraction and hence the resulting EOS will incorporate a mixed phase region (of varying pressure) regardless of whether one considers LCN or GCN.

\begin{figure*}
  \centering  \includegraphics[width=\textwidth]{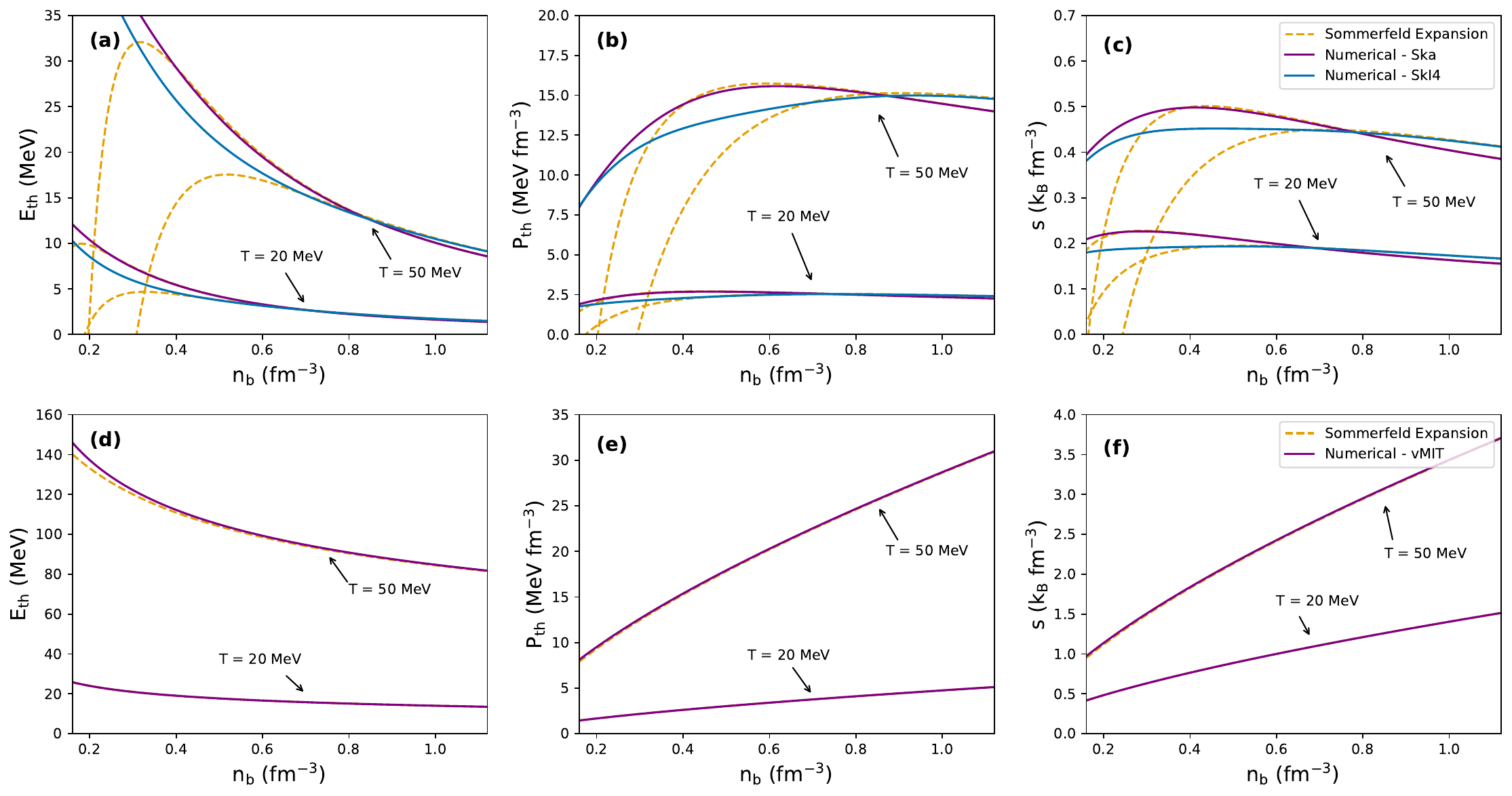}
  \caption{Comparing numerical and analytical calculations for the models employed in this work. Panels (a), (b) and (c) depict the results for the thermal energy per baryon, thermal pressure and entropy density for both hadronic models. Panels (d), (e) and (f) depict the same thermal properties but for the vector MIT bag model. In all panels, results are plotted for temperatures $T=20$ and 50 MeV. The numerical results are shown with the magenta and blue line solid lines, while the analytical approximations are plotted with the dashed orange curves. In panels (d), (e) and (f) the dashed curve is barely shown due to the fact that itcoincides almost perfectly with the solid curves.}
  \label{f1}
\end{figure*}

Let us now briefly discuss the derivation of hybrid EOSs for this very first stage of proto-neutron star evolution. The phase equilibrium for both the LCN and GCN cases, will be once again dictated by the demands of Eqs.~\eqref{eq35} and~\eqref{eq36}, respectively. However, due to the presence of neutrinos, an additional condition needs to be met. More precisely, the chemical potential of neutrinos in both phases needs to be equal. Then, in the case of LCN, the following system of equations needs to be solved, in order to determine the properties of the mixed phase for constant entropy per baryon and fixed lepton fraction~\cite{Pagliara-2010}
\begin{equation}
\begin{gathered}
    P^h = P^q, \hspace{0.5 cm} T^h=T^q \\
    \mu_b^h = \mu_b^q, \hspace{0.5 cm} \mu_\nu^h=\mu_\nu^q, \\
    n_p-n_e=0,\\
    2n_u-n_d-n_s-3n_e = 0,\\
    n_b=(1-\chi)n_b^h+\chi n_b^q,\\
    Y_ln_b=(1-\chi)n_e^h+\chi n_e^q+n_\nu,\\
    S_bn_b=(1-\chi)s_b^h+\chi s_b^q.
\end{gathered}
\end{equation}
In the scenario of GCN, the relevant system is modified as follows~\cite{Pagliara-2010}
\begin{equation}
\begin{gathered}
    P^h = P^q, \hspace{0.5 cm} T^h=T^q \\
    \mu_b^h = \mu_b^q,\hspace{0.5 cm} \mu_e^h=\mu_e^q, \hspace{0.5 cm} \mu_\nu^h=\mu_\nu^q, \\
   3 (1-\chi)n_p+\chi(2n_u-n_d-n_s)-3n_e=0,\\
    n_b=(1-\chi)n_b^h+\chi n_b^q,\\
    Y_ln_b=n_e+n_\nu,\\
    S_bn_b=(1-\chi)s_b^h+\chi s_b^q.
\end{gathered}
\end{equation}
Note that for values of pressure below the phase transition pressure (for which the systems above yield $\chi=0$), the hybrid EOS is described by the charge neutral hadronic phase with constant entropy per baryon and constant lepton fraction. Respectively, for values of pressure larger than the one at the end of mixed phase (marked by $\chi=1$) the hybrid model will be described by the charge neutral isentropic quark EOS with fixed lepton fraction.

\section{Results and Discussion} \label{section6}

Let us begin our analysis by discussing the construction of hybrid EOSs. In the present study, we have used two different widely employed parametrizations of the Skyrme Hamiltonian, namely SkI4~\cite{Reinhard-1995} and Ska~\cite{Kohler-1976}. Notably, SkI4 is softer compared to Ska leading to the prediction of lower radii for compact star configurations. For quark matter, we have employed two parametrizations of the vector MIT bag model. In particular, in the LCN case, for the strength of the repulsive interaction a value of $G_v=(g_v/m_V)^2=0.17$ fm$^{-2}$ was selected, while for the bag constant the value was set to $B^{1/4}=160$ MeV (vMIT(a) parametrization). In addition, in the GCN case, $G_v$ was set to 0.3 fm$^{-2}$ and $B^{1/4}$ to 165 MeV (vMIT(b) parametrization). These two parametrizations were selected so that the phase transition occurs for a similar pressure value in both LCN and GCN models for the cold case (using the same parametrization would yield significantly earlier phase transitions in the GCN scenario). Finally, for the description of the crust in cold hybrid stars we have used the widely employed Baym, Pethick and Sutherland EOS~\cite{Baym-1971}. For hot stellar configurations, following the approach of related studies, we have considered the presence of a crust, starting at a baryon density of 0.08 fm$^{-3}$, by utilizing the well-known Lattimer and Swesty EOS~\cite{Lattimer-1991}.

In the case of LCN, the derivation of the EOS for the second stage of proto-hybrid evolution, requires the derivation of the phase diagram, which in turn requires the calculation of multiple isothermal EOSs. The top panels of Fig.~\ref{f1} depict our numerical results on the thermal energy per nucleon, thermal pressure, and entropy density for hadrons and two different temperature values, $T=$20 and 50 MeV. In addition, we have plotted the predictions of the Sommerefeld expansion~\cite{Costantinou-2015b} to verify that our numerical code on the finite temperature extension of the Skyrme models works properly. As is evident, for both hadronic parametrizations, the agreement between numerical and analytical predictions is excellent in the asymptotic limit. At lower densities, the differences become very large, as higher order terms of the series expansion are required to achieve an accurate description of less degenerate hadronic matter. An interesting remark to be made is that the validity of the Sommerefeld expansion appears to be strongly dependant on the selected parametrization. More precisely, the range of validity for the expansion is significantly wider for the Ska model compared to the SkI4 model. For instance, considering the Ska EOS at $T=20$ MeV, the analytical expansion is sufficient for describing almost the entire range of densities related to a neutron star core, reproducing the numerical results for densities down to $\sim0.2$~fm$^{-3}$. In contrast, for the SkI4 EOS, the agreement between the numerical and analytical results begins for densities higher than $\sim0.4$~fm$^{-3}$. This difference between the two models is related to their corresponding proton fractions and proton effective masses. More precisely, the accuracy of the analytical approximation for the evaluation of any property describing the full system (both protons and neutrons) is determined by the accuracy of the approximation for the contribution from protons. As protons are significantly less degenerate in dense nuclear matter (at least when neutrinos are absent), the accuracy of the approximation is worse for their contributions. Now, the stiffer Ska EOS predicts larger proton fraction and therefore a larger proton density compared to the SkI4 (and hence higher Fermi momentum). In addition, Ska is found to predict a lower value for $m_p^*$ at a given density. At the end, $k_{Fp}^2/2m_p^*$ is larger for the Ska model and as a consequence protons can be thought as more degenerate for the same baryon density and temperature.

\begin{figure}
  \centering  \includegraphics[width=8.5 cm]{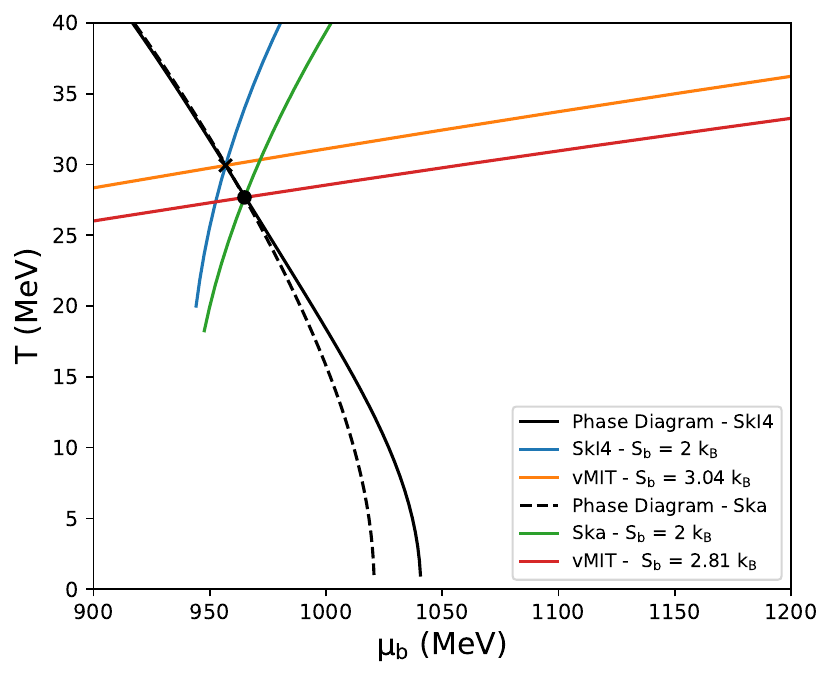}
  \caption{Phase diagram and isentropic curves in the $T-\mu_b$ plane. The black curves indicate the phase diagrams constructed by considering two different hadronic EOSs and one parametrization for the vector MIT bag model. The phase diagram derived using Ska is shown with the dashed black curve, while for SkI4 the black dashed solid curve was used. The green and blue lines correspond to constant entropy-per-baryon curves ($S_b=2$~$k_\mathrm{B}$) for the Ska and SkI4 hadronic models, respectively. The yellow and red curves denote isentropic trajectories for the quark model that cross the considered isentropic hadronic curves on each phase diagram.}
  \label{f2}
\end{figure}

In the bottom panels of Fig.~\ref{f1}, we present the numerical and analytical results for the thermal effects on quarks. As it is clear, both approaches yield nearly identical results for the entire density range that is potentially relevant for compact star applications (given that quark matter will necessarily appear beyond the saturation density). It is worth noting that the contributions of temperature in the context of the vector MIT bag model are completely independent of the selected parametrization (so we have plotted the results only for parametrization vMIT(a)). In particular, the thermal contribution for each thermodynamic variable is the same as it would be for a free Fermi-gas of quarks. This fact can be seen after some work on Eqs.~\eqref{eq23}~-~\eqref{eq25}, but it is more clearly reflected through the analytical expansions. In particular, the expansion (for q quarks) is dependant on terms of $m_q^*T/E_{Fq}$. However, due to the absence of scalar interactions we have $m_q^*=m_q$, and, therefore, it is obvious that the expansion terms are blind to the selected ($G_v,B$) parametrization for a given value of temperature and baryon density.

Let us proceed with the discussion on the construction of the hybrid EOS related to the second state of the proto-hybrid star evolution for the LCN case. As previously mentioned, during this period, it is thought that neutrinos have left the system. In addition, the entropy per baryon is considered to be constant and different for both phases. In Fig.~\ref{f2}, we present the phase diagrams derived using the aforementioned parametrizations for hadronic and quark matter. As is evident, the use of the Ska hadronic model leads to earlier phase transitions at low temperature values. As the temperature increases, going beyond 25 MeV, the phase diagrams are essentially identical despite the use of different hadronic EOSs. Furthermore, for the plotted isentropic hadronic EOSs one can observe that the softer EOS (SkI4) intersects the phase diagram at higher temperature and lower baryon chemical potential. Finally, the derived isentropic quark EOSs that intersect the phase diagram at the appropriate points, exhibit a significantly smoother variation of temperature with respect to the chemical potential compared to the corresponding hadronic models.

\begin{table}
\caption{Entropy per baryon at each phase and comparison of the numerical evaluation of the derivative
$\left(\frac{dP}{dT}\right)$ with the prediction from the right-hand side
(RHS) of the Clausius--Clapeyron equation for the two models considered (in units of fm$^{-3}$).}
\label{tab:cc_comparison}
\begin{ruledtabular}
\begin{tabular}{lcccc}
Model & $S_b^h$ ($k_\mathrm{B}$) & $S_b^q$ ($k_\mathrm{B}$) &
CC RHS $\left(\Delta S_b/\Delta v\right)$ &
$\left(dP/dT\right)^{\rm num}$ \\
\hline
SkI4 & 2 & 3.04 & $-0.243$ & $-0.243$ \\
Ska  & 2 & 2.81 & $-0.197$ & $-0.197$ \\
\end{tabular}
\end{ruledtabular}
\end{table}

In Table~\ref{tab:cc_comparison}, we report the difference in the entropies per baryon between the two phases. As is evident, the use of a stiffer model leads to lower differences in $S_b$. However, it is important to comment that in both cases (SkI4 and Ska) the entropy difference is not negligible, but of the order of $\sim1$. This yields a very different quark EOS, compared to the one that would be used at a first approximation under the assumption of equal entropies per baryon in both phases. To verify the thermodynamic consistency of our calculations related to this entropy per baryon difference, we have taken the phase diagram at the $P-T$ plane and evaluated its slope. In addition, we have evaluated the right hand side of the Clausius-Claperyon equation in Eq.~\eqref{cceq2} using the isentropic EOSs derived for both phases. In Table~\ref{tab:cc_comparison} we also present the relevant results showing that both methods for the evaluation of the $dP/dT$ slope lead to identical results. Interestingly, this demonstrates the self-consistency of the present calculations. 

\begin{figure*}[t]
  \centering  \includegraphics[width=\textwidth]{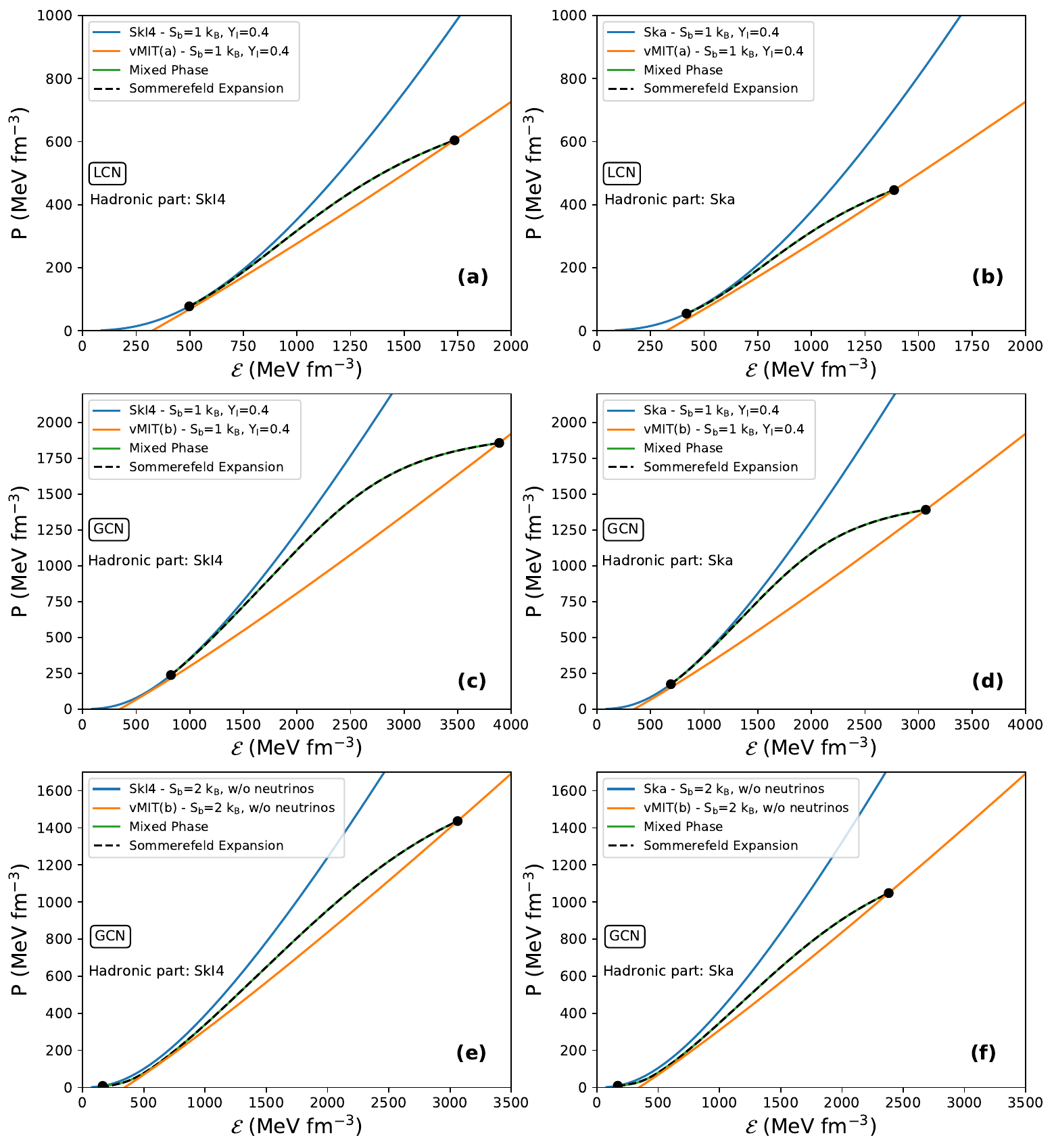}
  \caption{The hot hybrid EOSs constructed within the Gibbs framework in the present study. In all panels the blue curves denote the Skyrme hadronic EOS, the orange lines the quarks models, while the green ones the constructed mixed phases. In panels (a),(c) and (e) the SkI4 parametrization is used for the hadronic model, while in panels (b),(d) and (f) the Ska model was employed. Panels (a) and (b) indicate the EOSs for thermodynamic conditions related to first evolutionary stage under the assumption of LCN, while panels (c) and (d) contain the results for the same entropy per baryon and lepton fraction but for GCN. Panels (e) and (f) contain the hot EOSs described by constant entropy per baryon and zero neutrino density, charactering the second evolutionary stage of a proto-hybrid star. Finally, the black dashed curves, appearing in all panels, indicate the predictions of the Sommerefeld expansion~\cite{Costantinou-2015b} for the mixed phases (see the text for more details).}
  \label{f3}
\end{figure*}

For the construction of the hybrid EOS related to the first stage after the formation of proto-hybrid star we considered the global conservation of lepton fraction in the system. As a consequence, the phase transition is not abrupt but smooth, regardless of whether LCN or GCN is considered. In Figs.~\ref{f3}(a,b) we present the two hybrid models constructed using the two distinct hadronic EOSs in the LCN case.
As is evident, the use of the Ska parametrization leads to the deconfinement of quarks at lower density and pressure. However, the width of the mixed phase appears to be appreciably larger when a softer hadronic model is used. Figs.~\ref{f3}(c,d), show the EOSs for the same evolutionary stage but under the assumption of GCN. The effect of using different hadronic EOSs is the same as discussed above. However, there is a crucial difference in the width of the mixed phases. More precisely, mixed phase regions in the GCN scenario are considerably larger compared to the LCN case (the energy density range is more than double). As a consequence, it is not expected to find a pure quark matter core, in the earliest evolutionary stage when a GCN assumption is employed.

Figs.~\ref{f3}(e,f) indicate the EOSs that characterize a proto-hybrid star at the stage following the neutrino diffusion (stage 2). In contrast to the LCN case, here the appropriate EOS construction method is Gibbs, leading to an EOS with an extended mixed phase. Similarly to the case of the GCN model with trapped neutrinos (stage 1), the mixed phase is quite extensive (for SkI4 it ends after an energy of 3000 MeV fm$^{-3}$ is reached). Therefore, the existence of a purely quark matter core at this stage appears to be disfavoured as well.

To ensure that our mixed phase calculations are done properly, the resulting particle fraction and temperature values obtained for the mixed phase ranges were passed as an input in the Sommerfeld expansion~\cite{Costantinou-2015b} to yield the corresponding approximate results on pressure and energy density. These results, are plotted with dashed lines in all panels of Fig.~\ref{f3}. As is evident, the agreement between numerical and analytical calculations is remarkable. Interestingly, this also appears to be the case in Fig.~\ref{f3}(e,f) even though the mixed phase starts at very low density (one would expect a less degenerate system to break the approximation). However, since we are working with isentropic EOSs, the temperature value is also lower for low densities, which makes the approximation quite valid again.

\begin{figure*}[t]
  \centering  \includegraphics[width=\textwidth]{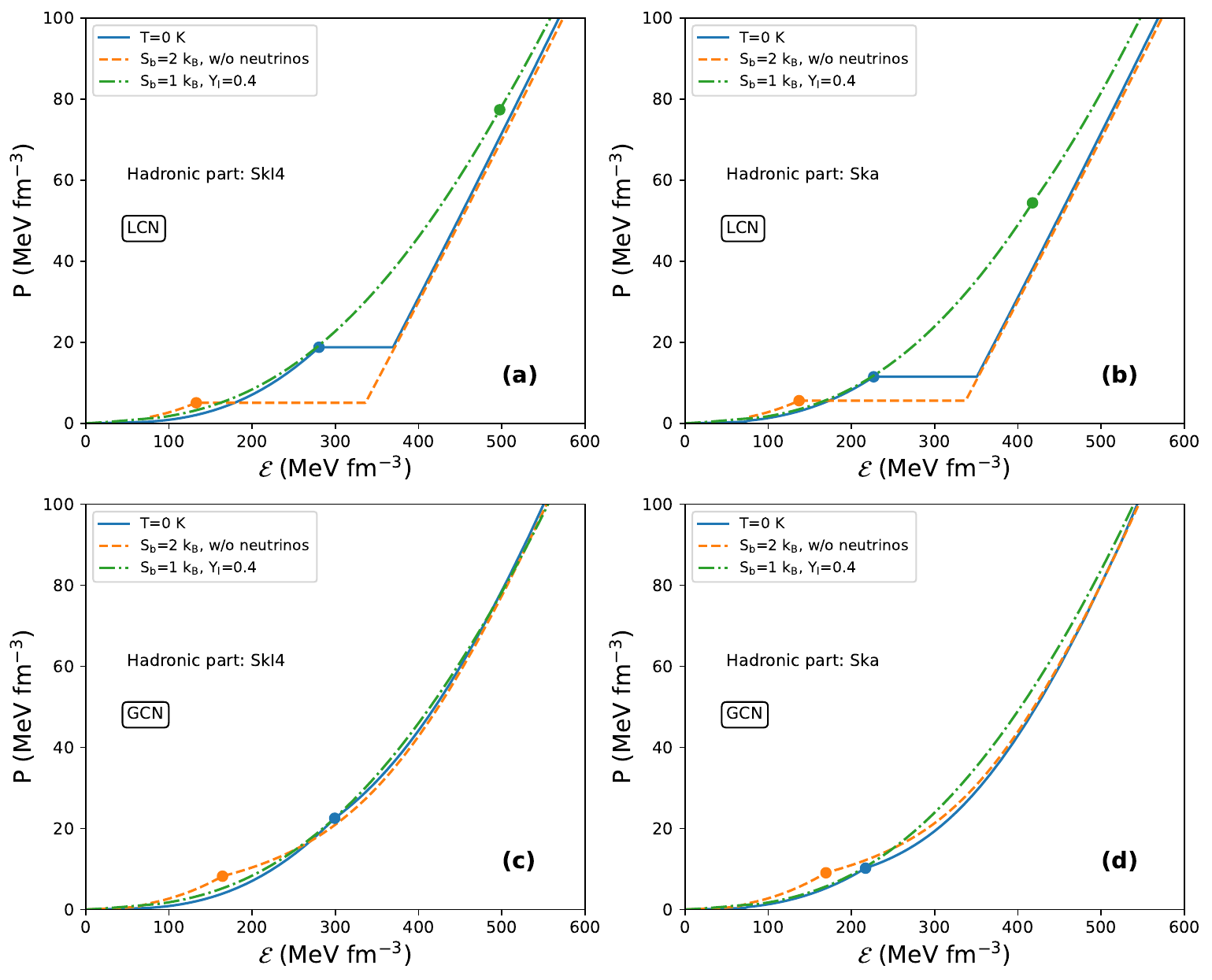}
  \caption{All of the hybrid EOSs constructed in this study. The solid blue curves denote the cold EOSs ($T=0$ K - final stage), the dashed orange curves denote the isentropic EOSs ($S_b=2$~$k_\mathrm{B}$ - second stage) without neutrino trapping (the entropy per baryon of 2 refers the hadronic part of the EOS. The relevant entropy per baryon in quark phase is that found in Table~\ref{tab:cc_comparison}), while the green dot-dashed curves indicate the isentropic EOSs with constant lepton fraction ($S_b=1$~$k_\mathrm{B}$ and $Y_l=0.4$ -  first stage). Panels (a) and (b) denote the results in the LCN case (local conservation of electric charge) for the SkI4 and Ska EOSs, respectively. Panels (c) and (d) denote the results in the GCN scenario (local conservation of electric charge) for the SkI4 and Ska hadronic models, respectively. The dot points indicate the onset of quark deconfinement (in the GCN case such points are beyond the axis limits for the green curves ).}
  \label{f4}
\end{figure*}

\begin{figure*}[t]
  \centering  \includegraphics[width=\textwidth]{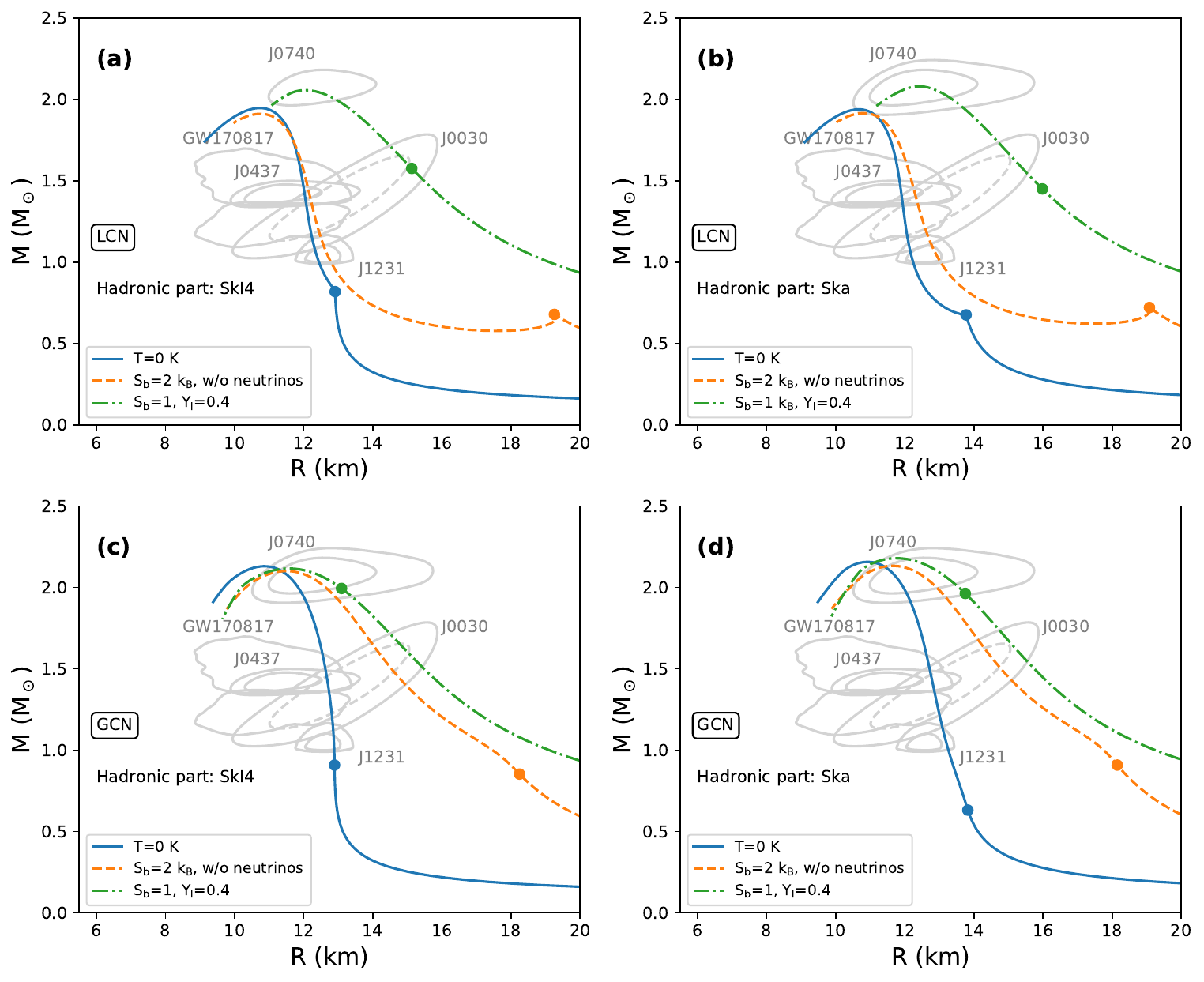}
  \caption{Mass-radius diagrams for the EOSs presented in Fig.~\ref{f4}. The solid blue curves denote the results for cold EOSs, the dashed orange curves denote the results for the isentropic EOSs ($S_b=2$~$k_\mathrm{B}$) without neutrino trapping (the entropy per baryon of 2 refers the hadronic part of the EOS. The relevant entropy per baryon in quark phase is that found in Table~\ref{tab:cc_comparison}), while the green dot-dashed curves indicate the case of isentropic EOSs with constant lepton fraction ($S_b=1$~$k_\mathrm{B}$ and $Y_l=0.4$) was used. In panels (a) and (c) the hadronic part of the EOS is modelled with the SkI4 parametrization, while in panels (b) and (c) Ska was used. The top panels indicate the LCN cases (vMIT(a) quark matter parametrization), while in the bottom panels GCN has been assumed (vMIT(b) quark matter parametrization). The dot points indicate the onset of quark deconfinement. The light gray curves indicate the measurements for the masses and radii of PSR J0030+0451~\cite{Vinciguerra-2024}, PSR J0740+6620,~\cite{Salmi-2024a},  PSR J0437-4715~\cite{Choudhury-2024},PSR J1231-1411~\cite{Salmi-2024} and GW170817~\cite{Abbott-2018}.}
  \label{f5}
\end{figure*}

\begin{figure*}[t]
  \centering  \includegraphics[width=\textwidth]{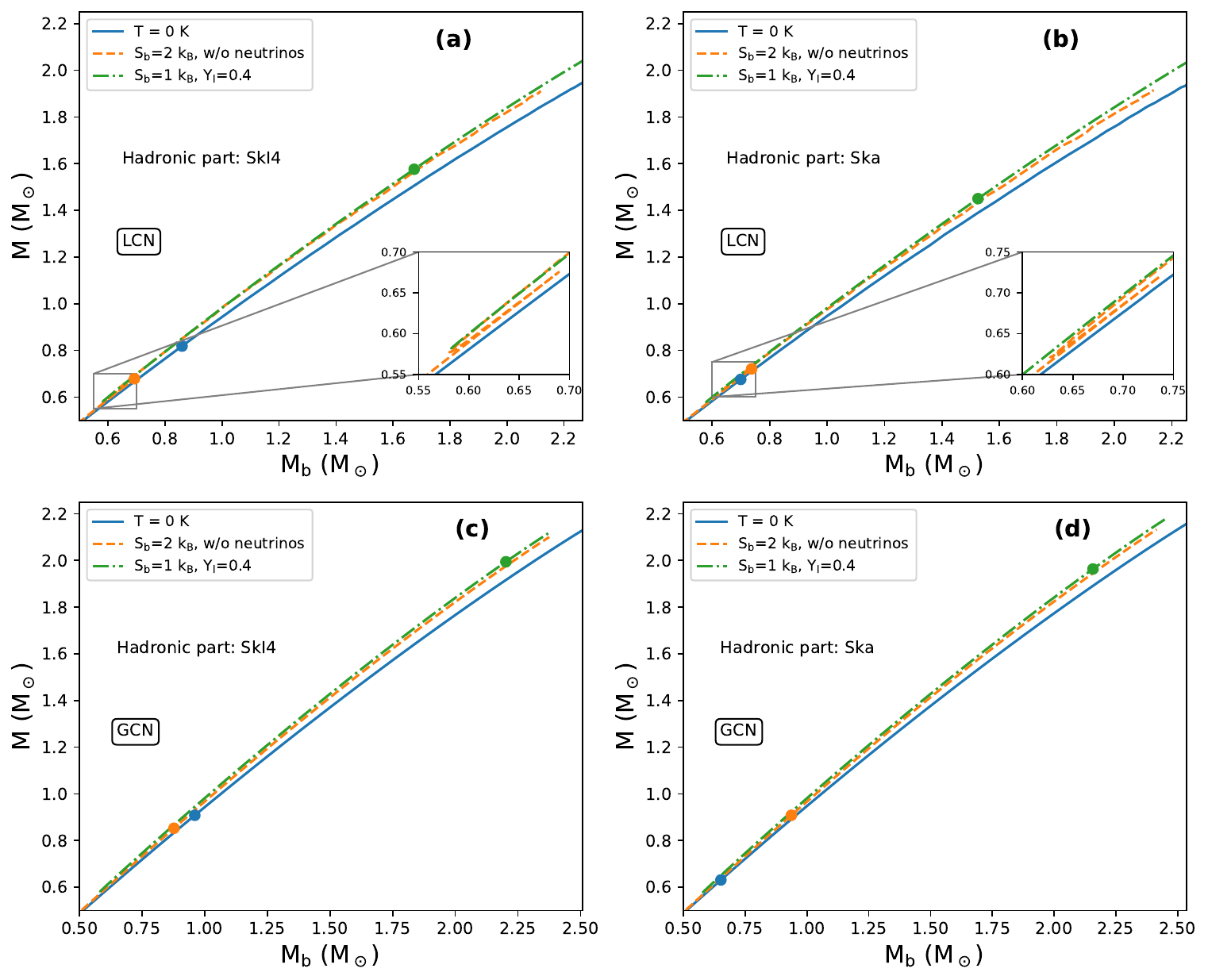}
  \caption{Gravitational mass as a function of the rest mass for the EOSs presented in Fig.~\ref{f4}. The solid blue curves denote the results for cold EOSs, the dashed orange curves denote the results for the isentropic EOSs ($S_b=2$~$k_\mathrm{B}$) without neutrino trapping (the entropy per baryon of 2 refers the hadronic part of the EOS. The relevant entropy per baryon in quark phase is that found in Table~\ref{tab:cc_comparison}), while the green dot-dashed curves indicate the case of isentropic EOSs with constant lepton fraction ($S_b=1$~$k_\mathrm{B}$ and $Y_l=0.4$) was used. In panels (a) and (c) the hadronic part of the EOS is modelled with the SkI4 parametrization, while in panels (b) and (c) Ska was used. The top panels indicate the LCN cases (vMIT(a) quark matter parametrization), while in the bottom panels GCN has been assumed (vMIT(b) quark matter parametrization). The dot points indicate the onset of quark deconfinement.}
  \label{f6}
\end{figure*}

Fig.~\ref{f4} depicts the hybrid EOSs related to each different evolutionary stage for proto-compact stars. An important remark is that during the formation period (high entropy per baryon and neutrino trapping) the pressure required for quark deconfinement is significantly larger compared to the following periods. In addition, during the second stage (high entropy per baryon and absence of neutrinos) the transition pressure is lower from the one for the cold case. The aforementioned results are common both in the LCN and the GCN cases. In the LCN scenario, as shown in Fig.~\ref{f4}(a,b), the EOS during the first stage is stiffer compared to the other two EOSs, for which the quark matter EOSs are very similar. While both the second stage EOS and the cold EOS suffer large density jumps, the width of the discontinuity is significantly larger in the former case. Considering Seidov's criterion~\cite{Seidon-1971,Bielich-2020} and the fact that this larger jump appears at relatively low pressure, one could expect a phase-transition-induced gravitational instability. As expected, in the GCN case, there is no discontinuity in the EOSs, regardless of the evolutionary stage. 

At the next step, we applied the derived hybrid EOSs to the well-known Tolman-Oppenheimer-Volkov (TOV) equations in order to study the potential structural evolution of proto-hybrid stars. In Fig.~\ref{f5}, we present the relevant results. Starting from the cold case, one can observe that the selected parametrizations are compatible to astronomical constraints for both the LCN and the GCN case. More precisely, the cold EOSs can effectively reproduce the masses and radii for PSR J0030+0451~\cite{Vinciguerra-2024}, PSR J0740+6620,~\cite{Salmi-2024a},  PSR J0437-4715~\cite{Choudhury-2024}, PSR J1231-1411~\cite{Salmi-2024} and GW170817~\cite{Abbott-2018}, at a $2\sigma$ level. For the finite temperature cases, one can observe that stellar configurations are significantly wider (higher radii) compared to the corresponding cold ones for low central pressures.

Interestingly, for the LCN scenario, shown in Figs.~\ref{f5}(a,b), the EOS for the first evolutionary stage predicts the larger maximum mass. This is of course related to the fact that, during that period, the phase transition induces an importantly less dramatic softening as a Gibbs construction scheme is used. The maximum mass for both the cold case and the intermediate evolution stage is rather similar, however slightly lower for the later. This was somewhat expected from our discussion of Fig.~\ref{f4} where we saw that while the quark matter EOS is rather similar in both of the aforementioned cases, the hybrid model related to second stage suffers a significantly larger density discontinuity. As a result of this large discontinuity, one can observe that the EOSs at hand leads to emergence of thermal twin star solutions (stars with identical mass and different radii). Note that the subject of thermal twins has recently regained attention with the relevant works of Refs.~\cite{Carlomagno-2024,Carlomagno-2024-twin}. 

In the case of GCN, presented in Figs.~\ref{f5}(c,d), all three EOSs now predict a similar value for the maximum gravitational mass. This difference from the LCN case may be attributed to the fact that all of the constructed EOSs have been derived using the Gibbs approach and therefore exhibit a moderate phase-transition-induced softening. Consequently, although the selected quark matter parametrization (vMIT(b)) predicts a phase transition at a stellar mass comparable to that of the LCN case (where vMIT(a) was used), it does not produce thermal twin star solutions.

At a first approximation, one can study the structural evolution of a proto-hybrid star as it cools down, using constant rest mass sequences~\cite{Bombaci-1996}. After its formation, a star conserves its total baryon number, meaning that its corresponding rest mass is fixed. The evaluation of this rest mass is given by the following expression~\cite{Lattimer-2001}
\begin{equation}
    M_B=m_b\int_0^R \frac{4\pi r^2n_b}{\sqrt{1-2m/r}}dr,
\end{equation}
where $m$ denotes the profile of the gravitational mass and $m_b$ is an effective nucleon mass often considered equal to 930.4 MeV~\cite{Lattimer-2001}. Using the aforementioned equation one can construct the relevant gravitational mass (or simply mass) - rest mass diagram. This is plotted in Fig.~\ref{f6}. We only depict stable stellar configurations and the x-axis limit is defined by the maximum gravitational (and in any case rest) mass for each cold model. A first observation that can be made is that the earlier the evolution stage, the larger the rest mass for a given gravitational mass. In addition, the zoomed regions indicate the appearance of thermal twin stars solutions for the second stage. As is evident, the instability triggered due to phase transition leads to a decrease in the rest mass with increasing central pressure. As a consequence, there are also thermal twin configurations in terms of the rest mass.

Probably the most important remark that one can make by observing Fig.~\ref{f6} is that the evolutionary history of a hybrid star appears to control the prediction for the maximum possible gravitational mass of the cold case. Specically, in the LCN case, the maximum rest mass during the intermediate evolution stage appears to be significantly lower compared the other two periods. That means that a star formed with a larger rest mass (compared to the maximum one for the second case), would turn unstable and potentially collapse as it evolves (and passes through the second stage), despite the fact that there are viable stellar configurations with the same mass at the final stage of evolution. For instance, let us assume that a star is formed at a rest mass equal to the one of maximum rest mass for the cold case. As is evident, while this star could in principle be formed, the configuration of equal rest mass during the second stage of evolution lies at gravitationally unstable regime. Therefore, while a cold EOS might predict a sufficiently high maximum mass to account for astronomical constraints, the evolution process might not allow for this configuration to ever exist. Therefore, the determination of whether a cold EOS can satisfy all current data is open until one can show that there are actual astrophysical paths that could lead the desired configurations~\cite{Strobel_2001,PhysRevD.90.064026,zy4h-6wj2}. A similar result was also found in the recent works of Refs.~\cite{Logoteta-2022,kunkel-2026}, in the context of different EOS construction methods (incorporating a mixed phase).

Notably, it should be commented that the story related to gravitational stability during the second stage of evolution may be different under certain assumptions. In the present study, the unstable region is mapped with the use of the well-known turning point criterion~\cite{Harrison-1965}. However, for proto-compact stars that are out of weak equilibrium the range of gravitational stability can be extended beyond the maximum mass configuration~\cite{Ghosh-2025,Canullan-Pascual-2025}. In addition, under the assumption of a {\it slow} phase transition (phase conversion timescale higher than oscillation timescale) the region of dynamical stability can also be widened~\cite{Pereira-2018,Rather-2024,Lugones-2023,Rau-2023,Mariani-2022,Laskos-2025}.

Finally, in the GCN case, the rest mass of the maximum mass configuration is the highest for the cold case. As a consequence, a star characterised by the rest mass of the maximum mass configuration for $T=0$ K would immediately collapse. As previously mentioned, such an effect has also been seen in the past for the evolution of hot hybrid stars incorporating a mixed phase~\cite{Logoteta-2022}. Interestingly, this is in contrast to the LCN case, where such a configuration could survive the first evolutionary stage without collapsing. 

Notably, since in the GCN case the EOSs display no discontinuity, one should not expect that an assumption of a {\it slow} phase conversion could extend the range of stability beyond the maximum mass configuration. More precisely, such an effect is directly related to a modification of the junction conditions at hadron-quark interface and therefore it is not present for continuous EOSs~\cite{Pereira-2018}. However, the consideration of out-of-equilibrium effects~\cite{Ghosh-2025,Canullan-Pascual-2025} could, in principle, extend the range of stability making the maximum mass configuration for the cold case a viable one. 

\section{Conclusion} \label{section7}
In the present study, we focused on the construction of hot isentropic EOSs for the modelling of hadron-quark phase transition in proto-neutron stars. For the description of the hadronic phase, two distinct Skyrme interactions (SkI4, Ska) were employed, while for quark matter the well-known vector MIT bag model was used. Furthermore, we considered the case of neutrino trapping characterising the very first stage of a proto-neutron star's evolution. In an attempt to provide a complete discussion on the possible hybrid configurations we worked under the assumptions of both LCN and GCN. 

To avoid the potential violation of the thermal equilibrium condition in LCN isentropic EOSs (without neutrinos), we allowed the two phases to be characterised by different entropies per baryon (derived with the use of the phase diagram). The thermodynamic consistency of the resulting models was verified by utilizing the Claussius-Clapeyron equation. Notably, the difference in the entropy per baryon between the hadronic and the quark phase was found to be larger when the softer SkI4 model was used. However, for both hadronic EOSs, the entropy difference was found to be relatively large, of order unity. The later means that the quark EOSs, along with the phase transition properties, could be significantly different from a scenario where a constant entropy per baryon has been imposed throughout the entire hybrid star. 

In the presence of neutrinos, the EOSs were constructed via the Gibbs approach, regardless of whether we considered LCN or GCN. In this scenario, the predicted mixed phases were found to be importantly more extensive in the GCN scenario. Furthermore, in both the LCN and GCN cases, the models that included neutrinos predicted significantly higher phase transition pressure compared to models of neutrino free matter. As a consequence, high pressure values may be required for quark deconfinement to occur during the first stage of proto-neutron star evolution.

With regard to the application of the aforementioned models for the study of hybrid star structure, we found that, in the LCN case, the maximum mass predicted by the EOS describing the first evolutionary stage was larger than the maximum mass predicted by the EOSs describing the subsequent stages. This was not the case in the GCN scenario, since all of the hybrid models were constructed within the Gibbs framework. 

At the next step, we calculated the rest mass of the constructed hybrid configurations in order to study their structural evolution. Crucially, it was found that, for all the models constructed in this study, there is no astrophysical path leading to the maximum-mass configurations predicted by the cold EOSs. This is of particularly important because, although a hybrid EOS may predict equilibrium stellar configurations with masses that account for astronomical observations, these configurations may never be found in nature if any preceding configuration along the evolutionary sequence is dynamically unstable.

An interesting future direction for the present research would involve the consideration of rotational effects on the constructed stellar configurations. In principle, newly formed compact stars could be rapidly rotating (uniformly or differentially) and this might have significant impact on their dynamical stability. In addition, as newly born compact stars are expected to vibrate and emit potentially detectable gravitational radiation, the derived EOSs could be applied to the study of non-radial stellar oscillations. Finally, studying the relevant dynamical stability of configurations beyond the maximum mass under the assumption of out-of-equilibrium effects or a {\it slow} phase conversion assumption would be critical.

\section*{ACKNOWLEDGMENTS}
P.L.-P. acknowledges that the research work was supported by the Hellenic Foundation for Research and Innovation (HFRI) under the 5th Call for HFRI PhD Fellowships (Fellowship No.~19175). This work is supported by the Croatian Science Foundation under the project Relativistic Nuclear Many-Body Theory in the Multimessenger Observation Era (HRZZ-IP-2022-10-7773). This paper was supported by the European Union – NextGenerationEU through the National Recovery and Resilience Plan 2021-2026 -- Institutional grant of University of Zagreb Faculty of Science (Nuclear Astrophysics). This research was supported by the European Union – NextGenerationEU through the National Recovery and Resilience Plan 2021–2026 Institutional grants of University of Zagreb Faculty of Science (PMF-PRESTIGE).

\bibliography{bibliography}

\end{document}